\documentclass[preprint,12pt]{elsarticle}

\usepackage{amssymb}
\usepackage{amsmath}
\usepackage{lineno}

\usepackage[dvipsnames]{xcolor}

\journal{Nuclear Inst. and Methods in Physics Research, A}

\begin{document}
%\linenumbers
\begin{frontmatter}



\title{Method to Reduce Noise for Measurement of {}$^7$Be and {}$^8$B Solar Neutrinos on Gallium-71}

\author[inst1]{J. Folkerts\fnref{jFolk}\corref{cor1}}
\fntext[jFolk]{\textit{Email:} jdfolkerts@shockers.wichita.edu \textit{Mailing address:} Jonathan Folkerts, Wichita State University, Campus box 32, 1845 Fairmount St., Wichita, KS 67260 \textit{Telephone:} (316) 978-3991}
\author[inst1]{N. Solomey\fnref{nSolo}}
\fntext[nSolo]{\textit{Email:} nick.solomey@wichita.edu}
\author[inst1]{B. Hartsock\fnref{bHart}}
\author[inst1]{T. Nolan\fnref{tNola}}
\author[inst1]{O. Pacheco \fnref{oPach}}
\author[inst2]{G. Pawloski\fnref{gPawl}}
\fntext[gPawl]{\textit{Email:} pawloski@umn.edu}
\cortext[cor1]{Corresponding Author}

%\author[inst1,inst2]{Author Three}

\affiliation[inst1]{organization={Wichita State University},
%Department and Organization
            addressline={1845 Fairmount St.}, 
            city={Wichita},
            postcode={67260}, 
            state={Kansas},
            country={United States of America}}

\affiliation[inst2]{organization={School of Physics and Astronomy, University of Minnesota Twin Cities},
%Department and Organization
%School of Physics and Astronomy, University of Minnesota Twin Cities, Minneapolis, Minnesota 55455, USA
            %addressline={100 Church St SE}, 
            city={Minneapolis},
            postcode={55455}, 
            state={Minnesota},
            country={United States of America}}

%\affiliation[inst4]{organization={University of Minnesota},
%%Department and Organization
%            addressline={2818 Como Avenue S.E.}, 
%            city={Minneapolis},
%            postcode={55414}, 
%            state={Minnesota},
%            country={United States of America}}

\begin{abstract}
Gallium solar neutrino experiments have historically used radiochemical counting to determine the event rate. A detector which directly measures the ejected electron and a nuclear de-excitation gamma could reduce background counting rates by way of a double-pulse technique. We find this reduction could be as large as 6 orders of magnitude in a 1.5 ton detector when compared with experiments that allow ground-state transitions. In our process, the detector measures the excited nuclear final state of the  germanium after an electron neutrino interacts with gallium nucleus through the charged-current interaction.  This results in a loss of approximately 90\% of the total neutrino signal, but higher energy processes are less suppressed. The neutrinos resulting from this higher energy selection are predominantly from the {}$^8$B and {}$^7$Be solar neutrino fluxes. 

\end{abstract}

%\includegraphics{grabs}%% 


\begin{highlights}
\item Gallium is increasingly available as a constituent of scintillating crystals which present promise in future solar neutrino detectors.
\item Requiring gallium neutrino interactions to produce an excited state filters the solar neutrino spectrum to heavily suppress most low-energy processes besides ${}^8$B and ${}^7$Be, which become the dominant processes.
\item The prompt de-excitation gamma from an excited state can produce a double electron-gamma signal, which could reduce noise from backgrounds such as ${}^{235}$U by as many as 6 orders of magnitude in a 100 ton detector.

\end{highlights}

\begin{keyword}
neutrino \sep gallium \sep double-pulse \sep solar \sep background reduction
%% Might want to get rid of these PACS is no longer maintained, and MSC is for maths
%%\PACS 0000 \sep 1111
%%\MSC 0000 \sep 1111
\end{keyword}

\end{frontmatter}

%\linenumbers

\section{Gallium Double Pulsing}\label{sec:GaDoublePulse}
The interaction of gallium with neutrinos has been of scientific interest since the early 1990's with the GALLEX experiment and the SAGE experiment \cite{ref:GALLEX,ref:SAGE}. These experiments used radiochemical methods to determine the number of neutrino interactions which had taken place. They were based on the charged-current weak interaction of an electron neutrino and gallium transmuting into ionized germanium and an electron, Equation \ref{eq:GalliumInteraction}. Traditional radiochemical experiments are not sensitive to excited states of end-state nucleii. Despite this, a significant fraction of these nucleii will be in an excited nuclear state, often with very short half lives. If a detector is sensitive to both the ejected electron and to the prompt gamma ray emission, a new technique for reducing noise by this two-particle signal can be developed.

\begin{equation}\label{eq:GalliumInteraction}
    \nu_e+~_{31}^{71}\textrm{Ga}\to e^-+~_{32}^{71}\textrm{Ge}^{+*} \to e^-+~_{32}^{71}\textrm{Ge}^{+} + \gamma 
\end{equation}

The biggest advantage of such a system would be an opportunity to reduce the rate of uncorrelated noise. For example, carbon 14 provides a significant background to current and past solar neutrino experiments such as Borexino \cite{ref:BorexinoBackground}. If it were possible to reduce the rate of such backgrounds by looking for two energy pulses separated in time or distance, the signal to noise ratio could be greatly improved.
% there would be significant gains for the scientific merit of such studies. 
Because nuclear decay rates obey Poisson statistics, the probability of a secondary event with rate $R$ occurring in time window $T$ following a primary event is given by Equation \ref{eq:rateRedux},
\begin{equation}\label{eq:rateRedux}
    P = 1 -e^{-RT}\approx RT,
\end{equation}
for $T\ll 1/R$. Consider a 1.5 ton GAGG detector like one proposed by Huber 2023 for testing the gallium anomaly \cite{ref:HuberGAGGDetector}. Using the impurities expected in the constituents of GAGG \cite{ref:GAGGContamination}, we find a singles rate of $11.9$ Hz. Applying a constraint of requiring two such events within 250 ns, the background is reduced by a factor of $3\times10^{-6}$. The background after this constraint is reduced to 3.1 /day. 
%The Borexino experiment has several backgrounds which are on the order of 10 per day, such as their {}$^{210}$Bi background at $11.5\pm1.0$/day. If a constraint of requiring two such events within 1 $\mu$s could be applied to this rate, it would be possible to reduce the background by a factor of $1.2\times10^{-10}$, reducing such a background from $\sim10$/day to $420$/gigayear. Such a constraint on Borexino is likely impossible, but a detector designed around double pulsing might be able to achieve such a background reduction.

A concept for a solar orbiter searching for neutrinos is already looking into the use of gallium double-pulsing to reduce harsh backgrounds in space \cite{ref:MyNIMArticle}. Preliminary lab tests for a CubeSat demonstrator of this detector technology, consisting of a 28 mm $\times$ 28 mm $\times$ 14 mm active GAGG volume read out on SiPMs, show that a 100 $\mu$s window only fails to reject between 0.16\% and 0.00076\% of background event pairs depositing \textgreater150 keV based on timing alone. An experiment working with any isotope which has accessible excited nuclear states might be able to employ a similar technique. For example, the MOON experiment looked for the interaction of a neutrino on ${}^{100}$Mo and the decay of ${}^{100}$Tc with a half life of 16 s \cite{ref:MOON}. Molybdenum has two excitation levels at 201 and 244 keV with half-lives $\sim5$ $\mu$s which might allow a review of the dataset to observe the excited state transitions and narrow the time window for background events.
%Any earth-based experiment with historical data that includes timing and energy of pulses could look for nuclear de-excitation gammas corresponding to their target which occur after their candidate neutrino event, especially in the higher energy regions of the solar neutrino spectrum using this or a similar double-pulse method. 
The five lowest energy states of germanium are shown in Table \ref{tab:NeutrinoReactions} \cite{ref:A=71}.

\begin{table}[htp!]
\begin{center}
\begin{tabular}{ | c | c | c | c | }
\hline
{Reaction Products} & {Energy Threshold} & {Photon Energy} & {Half Life} \\ 
 \hline\hline
 ${_{32}^{71}}$Ge$+e^-$& 232.5 keV & n/a & $11.43\pm0.03$ d \\  
 \hline
 ${_{32}^{71}}$Ge$+e^- + \gamma$& 407.4 keV & 175.0 keV & $81\pm 3$ ns \\  
 \hline
 ${_{32}^{71}}$Ge$+e^- + \gamma$& 430.9 keV & 198.4 keV & $20.2\pm0.12$ ms \\  
 \hline
 ${_{32}^{71}}$Ge$+e^- + \gamma$& 732.5 keV & 499.9 keV & not measured \\  
 \hline
 ${_{32}^{71}}$Ge$+e^- + \gamma$& 757.6 keV & 525.1 keV & not measured \\  
 \hline
\end{tabular}
\caption[Table of lowest energy states from the interaction of electron-type neutrinos with {}$^{71}$Ga. Shown are the energy threshold of the neutrino needed to produce the final state, the resulting energy of the de-excitation photon, and the half-life of the Ge state.]{Table of lowest energy states from the interaction of neutrinos with {}$^{71}$Ga.  Shown are the energy threshold of the neutrino needed to produce the final state, the resulting energy of the de-excitation photon, and the half-life of the Ge state. The 20.2 ms state may be problematic for use in background rejection, but it is very unlikely and can safely be removed from analyses.}
\label{tab:NeutrinoReactions}
\end{center}
\end{table}

%% NEW FROM REVIEW 1
Recent developments in gallium-based scintillators have given rise to potential targets which already contain gallium and are very fast. Cerium-doped Gadolinium Aluminum Gallium Garnet (GAGG or GAGG:Ce) is very fast with an 88 ns fast decay time and high light yield of 46,000 photons/MeV\cite{ref:GAGG}. This crystal contains about 23\% gallium by mass. A triggerd gamma ray source using a small GAGG volume and an external trigger detector was able to achieve detector efficiency between 78\% and 93\% on  various isotopes \cite{ref:doty2024gammaraydetectionefficiency}, and we expect that a large GAGG detector looking for solar neutrinos can achieve similar performance.

\begin{figure}[htbp]
    \centering
    \includegraphics[width=0.5 \textwidth]{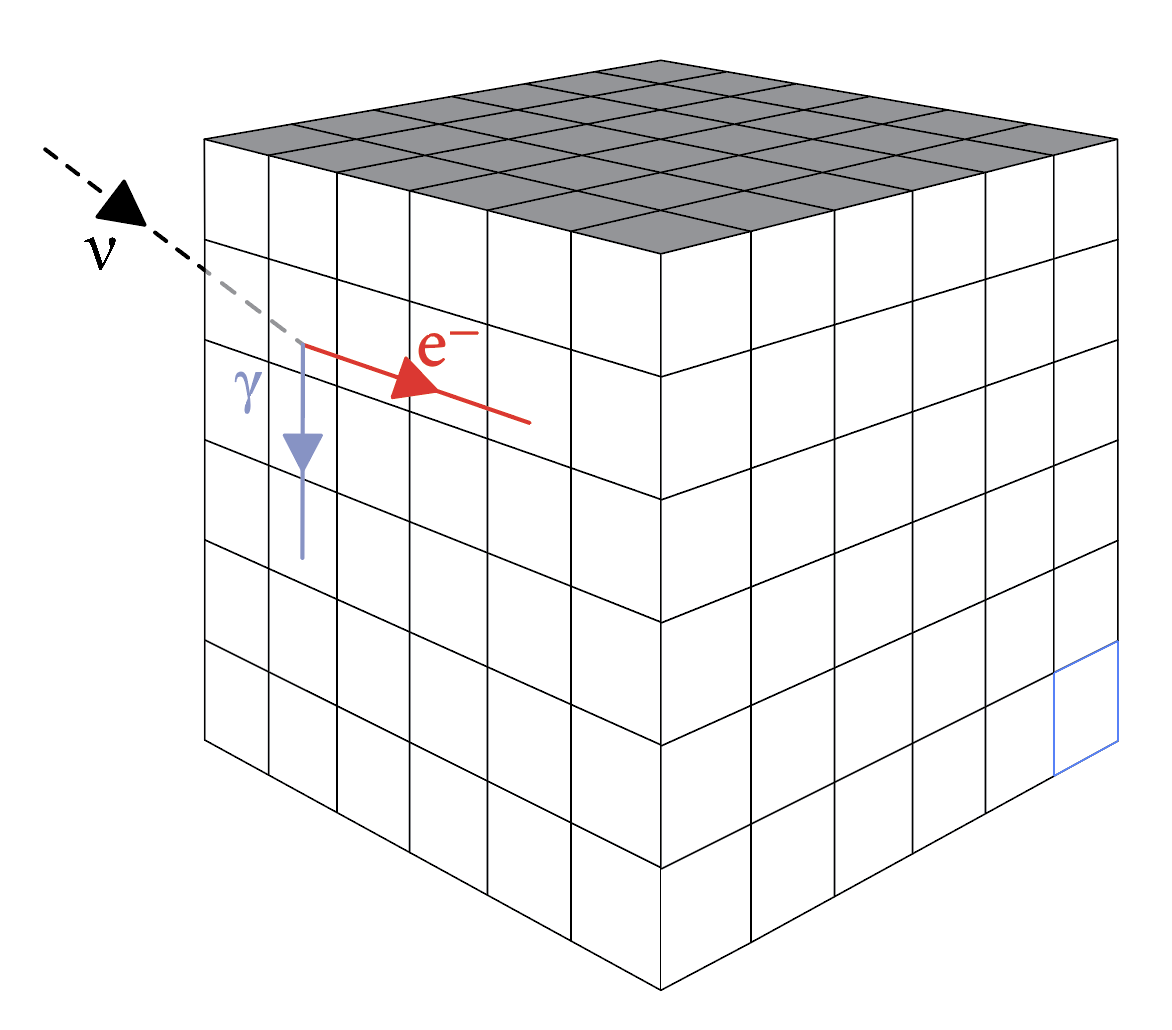}
    \caption[Diagram for a concept of a highly-voxelated neutrino detector with incident neutrino and isotropic electron and de-excitation gamma from the resulting nucleus.]{Diagram for a concept of a highly-voxelated neutrino detector with incident neutrino and isotropic electron and de-excitation gamma from the resulting nucleus.}
    \label{fig:voxelSketch}
\end{figure}

Recent research has also shown that $\beta$-Ga$_2$O$_3$ could be a very fast scintillator with decay times as fast as 12 ns and light yields as high as 6400 photons/MeV\cite{ref:Ga2O3}. This scintillator, if it could be produced in large quantities, would be 74\% gallium by mass. Either of these scintillators could allow for very fast timing resolution. In addition, if the detector was voxelated, such as in detectors like NO$\nu$A or Figure \ref{fig:voxelSketch}, it might be possible to discriminate the electron-gamma double signal by looking for the signals in two different volumes. Lab tests with 7 mm GAGG cubes bonded to SiPMs show that energy resolutions as good as $8.46\pm0.45$ @ ${}^{137}$Cs and timings as fast as $52.2\pm7.4$ ns are achievable. Initial tests also show that signals with as low as 81 keV could be separated from noise in GAGG \cite{ref:brooksThesis}. This gives us confidence in a gallium detector that does not need to use radiochemical methods to detect the neutrino interaction.

A GAGG detector operating at earth could measure 2.8 neutrinos /ton /year or a $\beta$-Ga$_2$O$_3$ could measure 9.7 neutrinos /ton /year.

\begin{figure}[htbp]
    \centering
    \includegraphics[width=0.65 \textwidth]{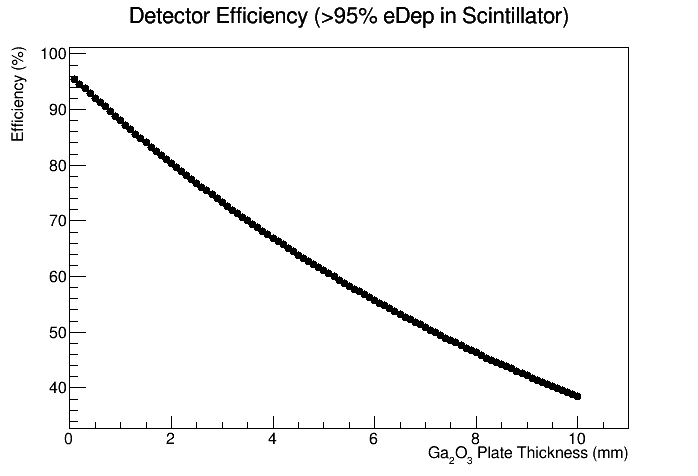}
    \caption[Fraction of 175 keV gamma rays with at least 95\% of their initial energy vs $\beta$-Ga$_2$O$_3$ thickness of crystal plate traversed.]{Fraction of 175 keV gamma rays with at least 95\% of their initial energy vs $\beta$-Ga$_2$O$_3$ thickness of crystal plate traversed.}
    \label{fig:Ga2O3Eff}
\end{figure}

\begin{figure}[htbp]
    \centering
    \includegraphics[width=0.65 \textwidth]{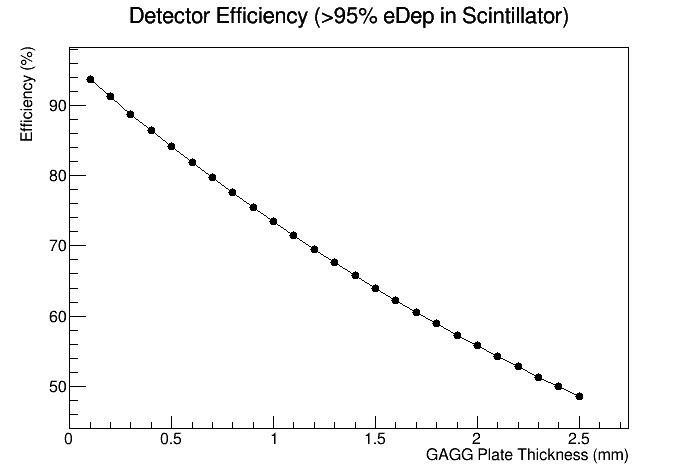}
    \caption[Fraction of 175 keV gamma rays with at least 95\% of their initial energy vs GAGG thickness of crystal plate traversed.]{Fraction of 175 keV gamma rays with at least 95\% of their initial energy vs GAGG thickness of crystal plate traversed.}
    \label{fig:GAGGEff}
\end{figure}

\begin{figure}[htbp]
    \centering
    \includegraphics[width=0.45 \textwidth]{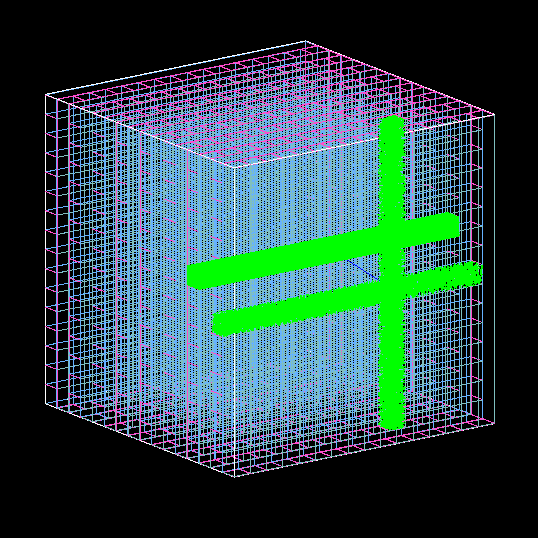}
    \caption[Simulated neutrino interaction in a detector made of $\beta$-Ga$_2$O$_3$ 1 $\times$ 1 $\times$ 16 mm voxels. The initial electron (8 MeV) crosses between two voxels, causing scintillation light (green lines). The secondary gamma ray (blue line) travels further into the detector before depositing in an inner voxel, causing more scintillation light.]{Simulated neutrino interaction in a detector made of $\beta$-Ga$_2$O$_3$ 1 $\times$ 1 $\times$ 16 mm voxels. The initial electron (8 MeV) crosses between two voxels, causing scintillation light (green lines). The secondary gamma ray (blue line) travels further into the detector before depositing in an inner voxel, causing more scintillation light.}
    \label{fig:Ga2O3SolarNeutrino}
\end{figure}

\begin{figure}[htbp]
    \centering
    \includegraphics[width=0.45 \textwidth]{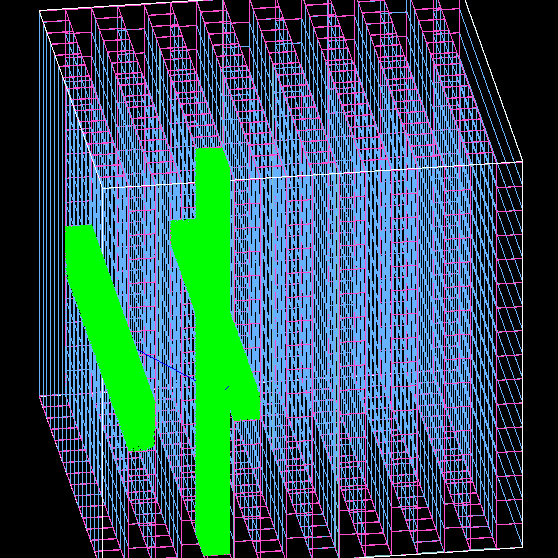}
    \caption[Simulated neutrino interaction in a detector made of GAGG 1 $\times$ 1 $\times$ 16 mm voxels. The initial electron (8 MeV) causes scintillation light (green lines) in two perpendicular voxels near the center. The secondary gamma (blue line) ray travels through the GAGG and deposits energy in two voxels near the edge, causing more scintillation light.]{Simulated neutrino interaction in a detector made of GAGG 1 $\times$ 1 $\times$ 16 mm voxels. The initial electron (8 MeV) causes scintillation light (green lines) in two perpendicular voxels near the center. The secondary gamma (blue line) ray travels through the GAGG and deposits energy in two voxels near the edge, causing more scintillation light.}
    \label{fig:GAGGSolarNeutrino}
\end{figure}

Using Geant4, we performed some initial studies of the 175 keV de-excitation gamma inside both GAGG and $\beta$-Ga$_2$O$_3$. The figure of interest from these simulations was the proportion of the gammas that had at least 95\% of their energy when traveling through the crystal. We chose this value based on the energy resolution from small test crystals \cite{ref:brooksThesis}. From Figures \ref{fig:Ga2O3Eff} and \ref{fig:GAGGEff}, we can see that the segments of a highly-voxelated design should have crystals no larger than $\sim$4 mm in $\beta$-Ga$_2$O$_3$ or $\sim$1 mm in GAGG for a $\sim$70\% gamma efficiency. Another Geant4 simulation with long square prisms as the detection volume is shown in Figures \ref{fig:Ga2O3SolarNeutrino} and \ref{fig:GAGGSolarNeutrino} with a solar-energy electron in one voxel with a 175 keV gamma ray escaping to interact in another. A detector like this could be constructed of many small modules and optimized for the particular crystal chosen.


%% END NEW FROM REVIEW 1

\section{Relative Amplification of ${}^7$Be and ${}^8$B Solar Neutrino Signals}\label{sec:7Be8BAmp}
Previous gallium neutrino experiments have used the Bahcall cross section for gallium interacting with solar neutrinos or one of several follow-up papers \cite{ref:Bahcall_CrossSec,ref:GalliumCrossSec3,ref:GalliumCrossSec2,ref:GalliumCrossSec4}. The original Bahcall method finds the cross section for the ground state and corrections for several of the common excited states. Of interest to this paper specifically is the behavior at these excited states of gallium.

\begin{figure}[htbp]
    \centering
    \includegraphics[width=\textwidth]{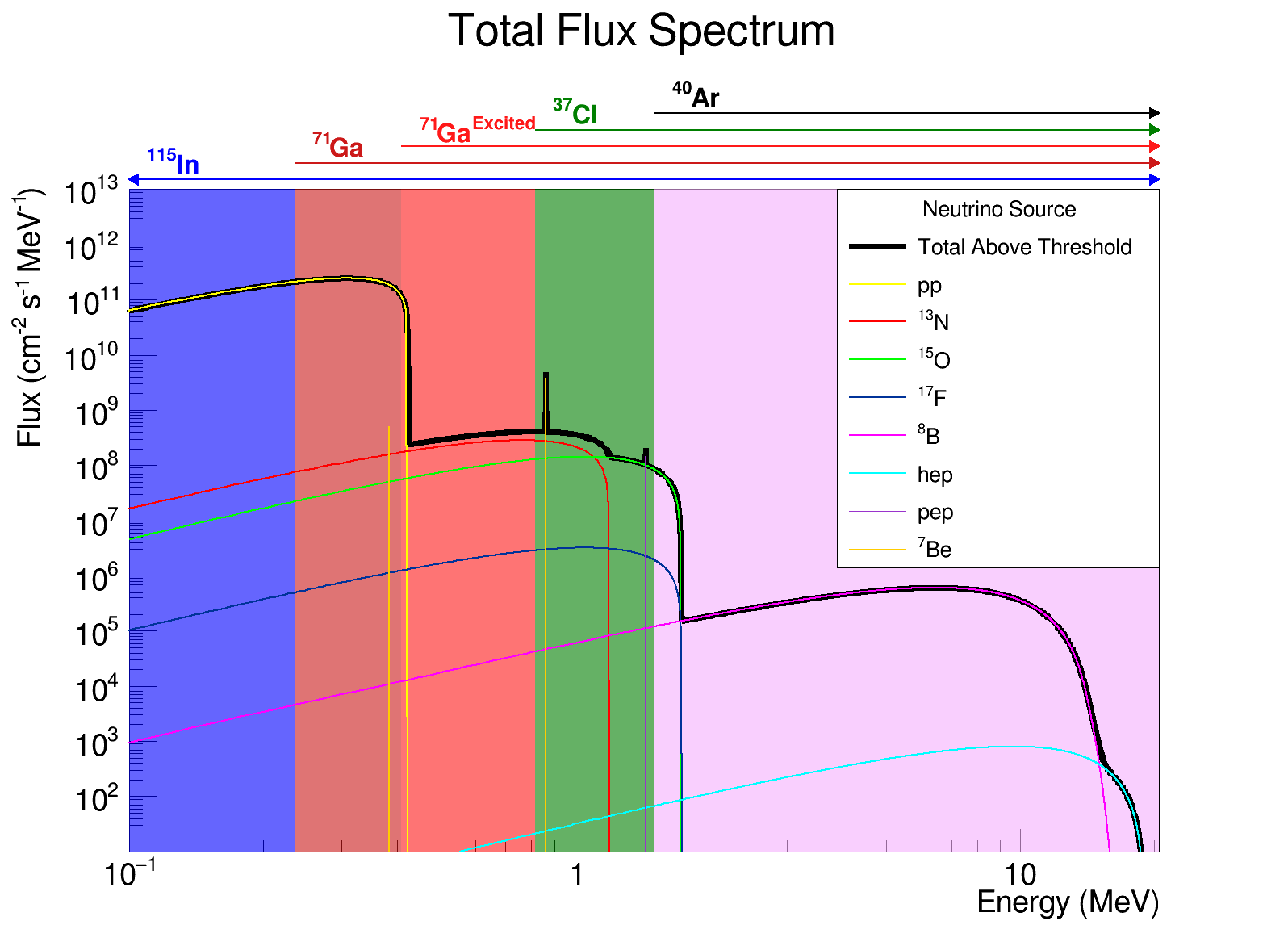}
    \caption[Solar Neutrino Spectrum produced using data from John Bahcall's standard solar model cite{ref:Bahcall_2005}. Each line is shown along with the total flux spectrum. Monoenergetic peaks have units of cm$^{-2}$ s$^{-1}$. Several elements and their thresholds are drawn and shaded on the graph.]{Solar Neutrino Spectrum produced using data from John Bahcall's standar solar model \cite{ref:Bahcall_2005}. Each line is shown along with the total flux spectrum. Monoenergetic peaks have units of cm$^{-2}$ s$^{-1}$. Several elements and their energy thresholds are drawn and shaded on the graph.}
    \label{fig:neutrinoSpectrum}
\end{figure}

For the rest of the paper, there are two definitions which are important. The effective cross section for a given neutrino process, $\sigma_i$, is an integral of the multiplication of the neutrino cross section, $\sigma(E)$ with the normalized flux vs energy, $\phi_i(E)$. A graph of the solar neutrino fluxes due to the various fusion processes can be seen in Figure \ref{fig:neutrinoSpectrum} \cite{ref:Bahcall_2005}. For a general solar neutrino, we define the effective cross section as the flux-weighted average of the effective cross section for all processes, i.e. $\sum_i\Phi_i\sigma_i/\sum_i\Phi_i$.

\begin{table}[htp!]
\begin{center}
\begin{tabular}{ | c | c | c | c | }
\hline
{Neutrino Process} & {Cross Section}  & {Excited Only Cross Section} & {Ratio}\\ 
 \hline\hline
\textit{pp} & 11.6 & 4.93$\times10^{-3}$&  0.0425\%\\  
 \hline
{}$^7$Be - 1 & 77.4 & 4.55& 5.88\% \\  
 \hline
{}$^7$Be - 2 & 22.9 & 0 & 0\%\\  
 \hline
{}$^{13}$N & 60.4 & 3.62 & 6.00\%\\  
 \hline
{}$^{15}$O & 114 & 15.9 & 14.0\%\\  
 \hline
\textit{pep} & 204 & 37.2 & 18.3\%\\  
 \hline
{}$^{8}$B & 2.40$\times10^4$ & 2.12$\times10^4$ & 88.3\%\\  
 \hline
{}$^{17}$F & 114 & 16.1 & 14.0\%\\  
 \hline
\textit{hep} & 7.14$\times10^4$ & 6.61$\times10^4$ & 92.3\%\\  
 \hline
\end{tabular}
\caption[Table of neutrino's generating process, the effective cross section of that process, the effective cross section of going only to an excited state of germanium, and the ratio between the two cross sections.. Cross sections are the effective cross section of the transmutation of $\nu_e +{}^{71}$Ga$\to e^-+{}^{71}$Ge$^*$ and are given in units of $10^{-46}$ cm$^{2}$.]{Table of neutrino's generating process, the effective cross section of that process, the effective cross section of going only to an excited state of germanium, and the ratio between the two cross sections.. Cross sections are the effective cross section of the transmutation of $\nu_e +{}^{71}$Ga$\to e^-+{}^{71}$Ge$^*$ and are given in units of $10^{-46}$ cm$^{2}$.}
\label{tab:crossSecSummary1}
\end{center}
\end{table}

Using a copy of the code that Bahcall used to calculate the cross sections for solar neutrinos, determining the effect of looking for excited states is very straightforward. More modern versions of the code exist, but they change very little that is relevant to this paper, they are built atop John Bahcall's original FORTRAN code, and the cross section has not changed significantly since Bahcall published the result of this code in 1997. The output files for each fusion source of solar neutrinos - \textit{pp}, \textit{pep}, ${}^8$B, etc - contain the contribution to the total cross section separated by end-state excitation level. To determine the effect of ignoring the ground state, the analysis was as simple as not adding their contribution to the effective cross section. Results from this analysis can be found in Tables \ref{tab:crossSecSummary1} and \ref{tab:crossSecSummary2}.

\begin{table}[htp!]
\begin{center}
\begin{tabular}{ | c | c | c | c | }
\hline
{Neutrino Process} & {$\sigma_\text{total}$} & {$\sigma_\text{Excited States}$} & {Ratio (\%)} \\ 
 \hline\hline
\textit{pp} & 10.7 & 4.5$\times10^{-3}$ & 4.2$\times10^{-2}$ \\  
 \hline
{}$^7$Be - 1 & 5.1 & 3.0$\times10^{-1}$ & 5.9 \\  
 \hline
{}$^7$Be - 2 & 1.7$\times10^{-1}$ & 0 & 0\\  
 \hline
{}$^{13}$N & 1.9$\times10^{-1}$ & 1.1$\times10^{-2}$ & 6.0 \\  
 \hline
{}$^{15}$O & 2.54$\times10^{-1}$ & 3.5$\times10^{-2}$ & 14.0 \\  
 \hline
\textit{pep} & 4.49$\times10^{-1}$ & 8.20$\times10^{-2}$ & 18.3 \\  
 \hline
{}$^{8}$B & 1.66 & 1.47 & 88.3 \\  
 \hline
{}$^{17}$F & 5.75$\times10^{-3}$ & 8.07$\times10^{-4}$ & 14.0 \\  
 \hline
\textit{hep} & 8.92$\times10^{-3}$ & 8.26$\times10^{-3}$ & 92.6 \\  
 \hline\hline
\textbf{Total} & \textbf{18.5} & \textbf{1.91} & \textbf{10.3}\\  
 \hline
\end{tabular}
\caption[Table of neutrino's generating process, the flux-weighted contribution of that process to the effective cross section, the contribution to the cross section when not considering ground state interactions, and the ratio of excited states only to all states. Cross section contributions are given in units of $10^{-46}$ cm$^2$.]{Table of neutrino's generating process, the contribution of that process to the effective cross section, the contribution to the cross section when not considering ground state interactions, and the ratio of excited states only to all states. Cross section contributions are in terms of their flux-weighted cross section, and they are given in units of $10^{-46}$ cm$^2$.}
\label{tab:crossSecSummary2}
\end{center}
\end{table}

\begin{figure}[htbp]
    \centering
    \includegraphics[width=0.9\textwidth]{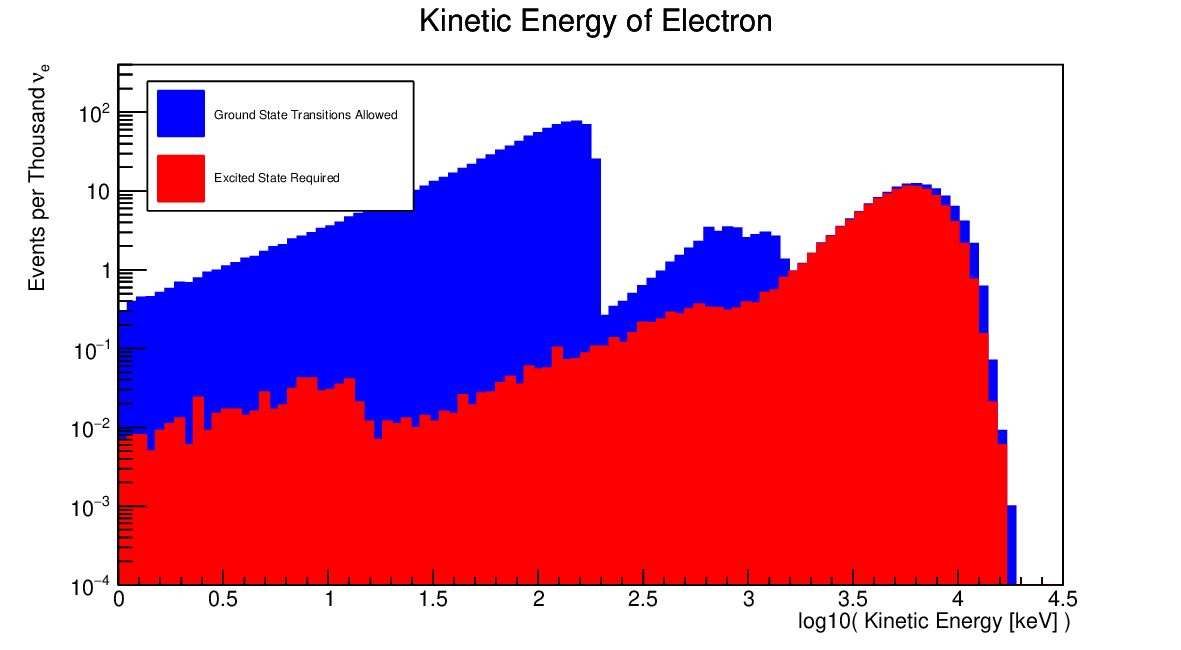}
    \caption[]{Histogram of the kinetic energy of outgoing electrons from the gallium neutrino interaction generated using MARLEY. The energy of all outgoing electrons is shown in blue, and the energy of the electrons associated with excited state transitions is shown in red.}
    \label{fig:outgoingElectrons}
\end{figure}

\begin{figure}[htbp]
    \centering
    \includegraphics[width=0.9\textwidth]{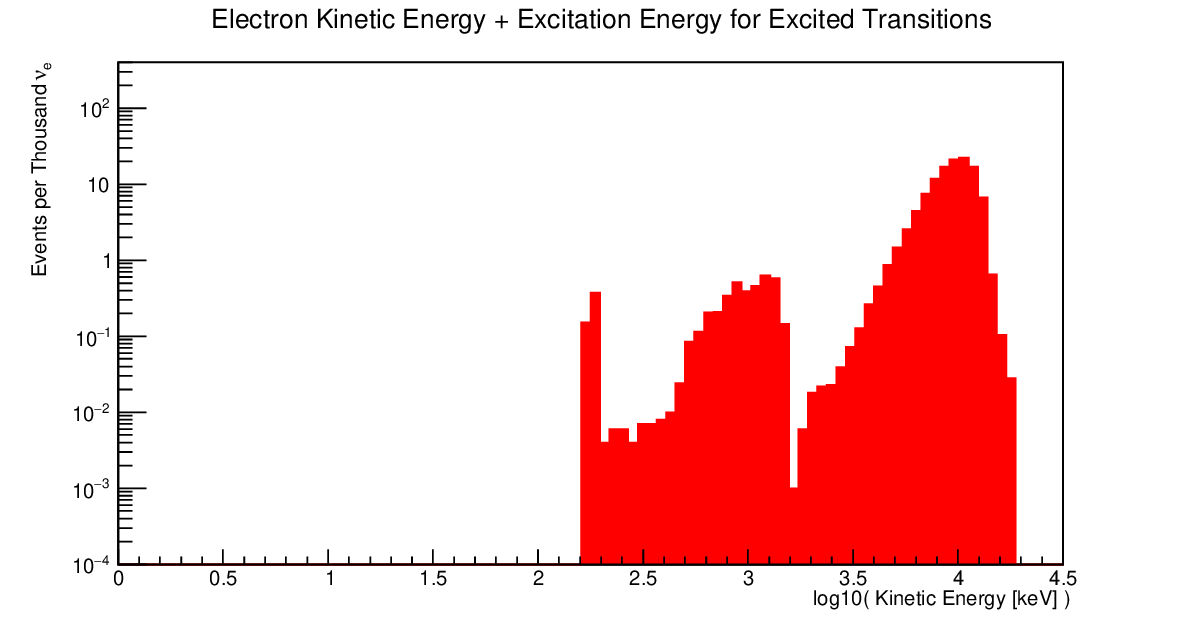}
    \caption[]{Histogram of the total energy of the outgoing electron and excited nucleus. Notice the separation of the {}$^8$B, {}$^7$Be + {}$^{15}$O + {}$^{13}$N, and \textit{pp} tail signals.}
    \label{fig:outgoingElectronsWithExcitation}
\end{figure}

Using the solar neutrino spectrum with the MARLEY event generator and the Bahcall Gamow-Teller factors, we find similar results \cite{ref:MARLEY}. A graph of the outgoing electron's kinetic energy can be seen in Figure \ref{fig:outgoingElectrons}. From this graph, we can see that we expect most excited state transitions to be associated with higher energy neutrinos than the ground state transitions. The most probable ground state interactions have $\approx 100$ keV of energy, but the excited states are most likely to be 1 MeV or more. We see that the {}$^8$B region is now the dominant energy region. Notice that each peak on the graph has been moved to the left by roughly the energy of the 1st excited state of gallium. These features are also slightly smeared by the appearance of other higher-energy excited states.
%that each of the features on the graph has moved slightly to the left. 
This corresponds to the energy being lost to the nucleus. The general effect is moving the feature down in energy, and spreading it out due to the multiple possible excitation levels in {}$^{71}$Ge. By adding the energy of the excitation level to the electron, the regions of the different processes separate noticeably, as can be seen in Figure \ref{fig:outgoingElectronsWithExcitation}. Combined with the energy resolution of the GAGG crystals, it should be possible to separate the signals by energy.

There are several interesting things to note from these results. First, when considering only excited states, the neutrino cross section averaged over the solar neutrino spectrum is 10\% of the total neutrino cross section. Second, the effective reduction in neutrino cross section is different across different neutrino processes owing to their differing energy profiles. Finally, in ground state transition measurements, the \textit{pp} neutrinos are the dominant neutrino measured for gallium, as can be seen in Table \ref{tab:crossSecSummary2}. When we consider only transitions to excited states of gallium, the \textit{hep} neutrinos have almost no reduction in rate, unlike most other neutrino processes, which can also be seen in Table \ref{tab:crossSecSummary2}. This process is not magnified enough to overcome its extremely low flux. However, by removing the ground state transitions, the {}$^8$B process becomes the dominant process.

\section{Noise Reduction with Gallium Double Pulses}
Any neutrino experiment that does not use radiochemical means of detecting neutrinos will suffer from some level of background due to longer-lived radioactive particles which were either present at the time of construction, or which are generated during operation from the passage of cosmic rays or cosmic ray showers. A double-pulse system can reduce the backgrounds of these particles by restricting interactions to very specific time windows.

To illustrate the background reduction gained by using double-pulsed signals in gallium, We will consider several elements that are primarily or entirely radioactive to show how much contamination a detector would be able to withstand. For a general radioactive background with half-life $T_{1/2}$ we can write Equation \ref{eq:probFalseDouble}, where $P$ is the overall probability of counting a time-correlated background as a signal, $P_R$ is the probability of rejecting a given single particle, and $P_t$, Equation \ref{eq:probTime}, is the probability of a second radioactive decay within the time window $t$ for $n$ radioactive particles. These equations assume that for an experimentally relevant timescale, the number of radioisotopes is not changing significantly.

\begin{equation}\label{eq:probFalseDouble}
    P = \left(1-P_R\right)^2\cdot P_t
\end{equation}
\begin{equation}\label{eq:probTime}
    P_t = 1 - 2 ^{-\frac{nt}{T_{1/2}}}
\end{equation}

This lets us calculate how much of a given radioactive isotope will begin to cause false double pulses unless $P_R$ approaches 1, given in Table \ref{tab:nAtomsProblem}. We have chosen a number of radioactive isotopes across a range of half-lives.  We include {}$^{71}$Ge and {}$^{152}$Gd because the former will be generated via neutrino interactions, and the latter has a small natural abundance. A GAGG detector will generate 13 atoms of germanium 71 per ton per year on earth due to neutrino interactions. GAGG is 64\% gadolinium by mass which corresponds to $4.9\times10^{21}$ atoms of {}$^{152}$Gd per kg of GAGG. A group working on the PIKACHU experiment has studied the impurity level of four radioactive isotopes in GAGG -- {}$^{238}$U, {}$^{235}$U, {}$^{232}$Th, and {}$^{40}$K \cite{ref:GAGGContamination}, and we have included these isotopes in the analysis.

\begin{table}[htp!]
\begin{center}
\begin{tabular}{ | c | c | c | c |}
\hline
{Isotope} & {Half-life} & $n:P_{t=\text{200 ns}}\approx1/2
$ & Expected Impurity\\ 
 \hline\hline
{{}$^{71}$Ge} & 11.43 d & 4.94$\times10^{12}$ &  $<1$ atom/ton\\  
 \hline 
{{}$^{152}$Gd} & $1.10\times10^{14}$ yr  &2.00$\times10^{28}$ & 1$\times10^{-3}$ g/g\\  
 \hline 
%{{}$^{14}$C} & 5730 yr & 9.04$\times10^{17}$ & $\lesssim2\times10^{-15}$ g/g\\ \hline
%{{}$^{115}$In} & $4.41\times10^{14}$ yr& 6.95$\times10^{28}$ \\   \hline
%{{}$^{239}$Pu} & $2.41\times10^{4}$ yr& $2.81\times10^{17}$ \\   \hline
{{}$^{235}$U} & $7.04\times10^{8}$ yr& $8.22\times10^{21}$ & $1.2\times10^{-10}$ g/g\\  
 \hline
{{}$^{238}$U} & $4.47\times10^{9}$ yr& $7.05\times10^{23}$ & $3.1\times10^{-9}$ g/g\\  
 \hline
{{}$^{}$Th} & $1.40\times10^{10}$ yr& $2.21\times10^{24}$ & $1.4\times10^{-9}$ g/g\\  
 \hline
{{}$^{40}$K} & $1.25\times10^{9}$ yr& $1.97\times10^{23}$ & $8.7\times10^-11$ g/g\\  
 \hline
%{{}$^{85}$Kr} & 10.8 yr & 1.70$\times10^{15}$\\   \hline
%{{}$^{210}$Bi} & 5.0 d & 2.16$\times10^{12}$ \\  \hline
%{{}$^{210}$Po} & 138.4 d & 5.68$\times10^{13}$ \\  \hline 
\end{tabular}
\caption[]{Table of several radioactive isotopes for comparison to how much is required to produce significant noise. For each isotope, the number of particles required for that isotope to have a 50\% chance of creating a double pulse inside a 200 ns time window and expected impurity are shown.}
\label{tab:nAtomsProblem}
\end{center}
\end{table}

The passage of muons through most materials can cause radioactive activation within a detector, and GAGG is no exception. Using the muon flux at sea level \cite{ref:muonFlux}, we simulated exposure of a 1 ton cube of GAGG to muons vertically incident. The simulation counted each radioactive isotope generated which had a half life of less than 10 years. A Table \ref{tab:GAGGMuonActivation} summarizes the results of this simulation. Tables \ref{tab:mu+TopParticles} and \ref{tab:mu-TopParticles} additionally show the one or two most abundant nuclei that result from $\mu^+$ and $\mu^-$ passage. From this table there are two regions of interest. If we make the extremely pessimistic assumption that all of the radioactive isotopes are produced at once, the long half-life region still shows that a GAGG detector looking for gallium double pulse signals is unlikely to have issues rejecting backgrounds due to the nuclear decays. The region with the short half-lives could present a significant problem to a GAGG detector. This detector would need a secondary method of rejection like an external veto or muon-tagging software. At sea level, muons interact with 1 ton of GAGG at a rate of $\sim200$ Hz. A veto that rejects a 100 $\mu$s window following a muon would therefore reject $\sim2$ \% of the total time.

\begin{table}[htbp]
    \centering
    \begin{tabular}{ | c | c | c | }
    \hline
Half Life Range & Events (/yr /ton) & {$n:P_t\approx1/2$}\\ 
 \hline\hline
$<1$ ns & $2.197(4)\times10^5$ & $2.5\times10^{-2}$ \\
 \hline
1 - 10 ns & $2.936(5)\times10^5$ & $2.5\times10^{-2}$ \\
 \hline
10 - 100 ns & $3.34(7)\times10^4$ & $2.5\times10^{-1}$ \\
 \hline
100 ns - 1000 ns & $2.46(2)\times10^4$ & $2.5\times10^{0}$ \\
 \hline
1 - 10 $\mu$s & $9.2(3)\times10^2$ & $2.5\times10^{1}$ \\
 \hline
10 - 100 $\mu$s & $3.62(6)\times10^3$ & $2.5\times10^{2}$ \\
 \hline
100 - 1000 $\mu$s & $2.5^{+1.4}_{-0.9}$ & $2.5\times10^{3}$ \\
 \hline
1 - 10 ms & $44.5^{+4.9}_{-4.5}$ & $2.5\times10^{4}$ \\
 \hline
10 - 100 ms & $3.24(2)\times10^4$ & $2.5\times10^{5}$ \\
 \hline
100 - 1000 ms & $3.17(4)\times10^3$ & $2.5\times10^{6}$ \\
 \hline
1 - 10 s & $6.019(6)\times10^{5}$ & $2.5\times10^{7}$ \\
 \hline
10 - 100 s & $3.022(1)\times10^6$ & $2.5\times10^{8}$ \\
 \hline
100 - 1000 s & $7.573(3)\times10^6$ & $2.5\times10^{9}$ \\
 \hline
1 - 10 ks & $1.3453(3)\times10^7$ & $2.5\times10^{10}$ \\
 \hline
10 - 100 ks & $1.0819(2)\times10^7$ & $2.5\times10^{11}$ \\
 \hline
100 - 1000 ks & $1.4978(9)\times10^6$ & $2.5\times10^{12}$ \\
 \hline
1 - 10 Ms & $1.1377(2)\times10^7$ & $2.5\times10^{13}$ \\
 \hline
10 - 100 Ms & $1.5550(9)\times10^7$ & $2.5\times10^{14}$ \\
 \hline
100 - 1000 Ms & $2.0188(3)\times10^7$ & $2.5\times10^{15}$ \\
 \hline
1 - 10 Gs & $2.886(3)\times10^6$ & $2.5\times10^{16}$ \\
 \hline
10 Gs - 10 yr& $<0.5$ & $2.5\times10^{17}$ \\
 \hline
\end{tabular}
    \caption{Table of events per year induced by muon flux at sea level incident on a 1 ton cube of GAGG for all byproducts with half-life $<10$ years and the number of such particles so that $P_t\approx\frac{1}{2}$. All entries between 1 and 100 ns are nuclear de-excitation gamma rays. Less than 1 per thousand entries above 1 second are nuclear de-excitation gamma rays. 13\% of entries in the 100 - 1000 ms range are de-excitation gammas. Entries below 1 ns are 15\% de-excitation gammas. The 2.5 number in the last column is a result of the chosen 200 ns time window and assuming all isotopes in the 1-10 ns range have 5 ns half lives on average, the 10-100 ns isotopes have 50 ns half lives on average, etc.}
    \label{tab:GAGGMuonActivation}
\end{table}

\begin{table}[htp!]
\centering
\footnotesize
\begin{tabular}{|c|c|c|c|}
\hline
Half Life Range & Particle & Count (/yr) & Excitation Level \\
\hline
$<1$ ns  & ${}^{153}$Gd & 7133.5 & 9 \\
 & ${}^{26}$Mg & 301 & 5 \\
\hline
1 - 10 ns & ${}^{156}$Gd & 56652.5 & 1 \\
 & ${}^{154}$Gd & 39252 & 1 \\
\hline
10 - 100 ns  & ${}^{159}$Gd & 21370.5 & 1 \\
 & ${}^{154}$Gd & 2093.5 & 2 \\
\hline
100 - 1000 ns  & ${}^{157}$Gd & 18934 & 1 \\
 & ${}^{6}$Li & 81.5 & 3 \\
\hline
1 - 10 $\mu$s & ${}^{153}$Gd & 300.5 & 2 \\
& ${}^{140}$Ce & 237 & 2 \\
\hline
10 - 100 $\mu$s & ${}^{157}$Gd & 1716.5 & 2 \\
& ${}^{153}$Gd & 1155 & 4 \\
\hline
100 - 1000 $\mu$s & n/a & n/a & n/a \\
\hline
1 - 10 ms & ${}^{138}$Ce & 32.5 & 1 \\
\hline
10 - 100 ms & ${}^{155}$Gd & 9571.5 & 3 \\
 & ${}^{24}$Na & 2.5 & 1 \\
\hline
100 - 1000 ms & ${}^{6}$He & 9571.5 & 1 \\
\hline
1 - 10 s & ${}^{25}$Na & 11.5 & 0 \\
& ${}^{16}$N & 1.5 & 0 \\
\hline
10 - 100 s & ${}^{139}$Ce & 1660.5 & 1 \\
 & ${}^{135}$Ce & 4.5 & 1 \\
\hline
100 - 1000 s & ${}^{161}$Gd & 4454.5 & 0 \\
& ${}^{15}$O & 4047.5 & 0 \\
\hline
1 - 10 ks & ${}^{70}$Ga & 17261.5 & 0 \\
& ${}^{68}$Ga & 9975.5 & 0 \\
\hline
10 - 100 ks & ${}^{159}$Gd & 46511 & 0 \\
& ${}^{72}$Ga & 2678 & 0 \\
\hline
100 - 1000 ks & ${}^{67}$Ga & 3120 & 0 \\
& ${}^{67}$Cu & 50.5 & 0 \\
\hline
1 - 10 Ms & ${}^{141}$Ce & 559 & 0 \\
& ${}^{7}$Be & 3.5 & 0 \\
\hline
10 - 100 Ms & ${}^{153}$Gd & 3166.5 & 0 \\
& ${}^{139}$Ce & 2745 & 0 \\
\hline
100 - 1000 Ms & ${}^{3}$H & 56.5 & 0 \\
 & ${}^{155}$Eu & 2.5 & 0 \\
\hline
\end{tabular}
\caption{Table of the most abundant particles generated by one year of $\mu^+$ fluence at sea level on 1 ton of GAGG. Up to two of the most common particles generated are shown per half life window.}
\label{tab:mu+TopParticles}
\end{table}

{
\begin{table}[htp!]
\centering
\footnotesize
\begin{tabular}{|c|c|c|c|}
\hline
Half Life Range & Particle & Count (/yr) & Excitation Level \\
\hline
$<1$ ns & ${}^{8}$Be & 178318 & 0 \\
 & ${}^{26}$Mg & 16794.5 & 9 \\
\hline
1 - 10 ns & ${}^{156}$Gd & 17690.5 & 1 \\
 & ${}^{154}$Gd & 11026.5 & 1 \\
\hline
10 - 100 ns & ${}^{159}$Gd & 5873 & 1 \\
 & ${}^{154}$Gd & 575.5 & 2 \\
\hline
100 - 1000 ns & ${}^{157}$Gd & 5502 & 1 \\
 & ${}^{6}$Li & 26 & 3 \\
\hline
1 - 10 $\mu$s & ${}^{140}$Ce & 100.5 & 2 \\
 & ${}^{153}$Gd & 83.5 & 2 \\
\hline
10 - 100 $\mu$s & ${}^{157}$Gd & 501.5 & 2 \\
 & ${}^{153}$Gd & 291.5 & 4 \\
\hline
100 - 1000 $\mu$s & ${}^{153}$Tb & 2.5 & 2 \\
\hline
1 - 10 ms & ${}^{138}$Ce  & 10.5 & 1 \\
 & ${}^{15}$B  & 1.5 & 0 \\
\hline
10 - 100 ms & ${}^{12}$B & 11548.5 & 0 \\
 & ${}^{13}$B & 7791 & 0 \\
\hline
100 - 1000 ms & ${}^{25}$Ne & 1975.5 & 0 \\
 & ${}^{64}$Co & 293.5 & 0 \\
\hline
1 - 10 s & ${}^{16}$N & 573259 & 0 \\
 & ${}^{26}$Na & 18482.5 & 0 \\
\hline
10 - 100 s & ${}^{160}$Eu & 2518342 & 0 \\
 & ${}^{68}$Cu & 254813.5 & 0 \\
\hline
100 - 1000 s & ${}^{71}$Zn & 2854643.5 & 0 \\
 & ${}^{27}$Mg & 3908269 & 0 \\
\hline
1 - 10 ks & ${}^{69}$Zn & 4042532 & 0 \\
 & ${}^{159}$Eu & 3585107.5 & 0 \\
\hline
10 - 100 ks & ${}^{157}$Eu & 7071068 & 0 \\
 & ${}^{159}$Gd & 1795430.5 & 0 \\
\hline
100 - 1000 ks & ${}^{67}$Cu & 513987 & 0 \\
 & ${}^{153}$Sm & 391700 & 0 \\
\hline
1 - 10 Ms & ${}^{156}$Eu & 6938411.5 & 0 \\
 & ${}^{149}$Eu & 2511851 & 0 \\
\hline
10 - 100 Ms & ${}^{65}$Zn & 1308802 & 0 \\
 & ${}^{153}$Gd & 183485 & 0 \\
\hline
100 - 1000 Ms & ${}^{155}$Eu & 9057999 & 0 \\
 & ${}^{154}$Eu & 6788553.5 & 0 \\
\hline
1 - 10 Gs & ${}^{150}$Eu & 2886105 & 0 \\
\hline
\end{tabular}
\caption{Table of the most abundant particles generated by one year of $\mu^-$ fluence at sea level on 1 ton of GAGG. Up to two of the most common particles generated are shown per half life window.}
\label{tab:mu-TopParticles}
\end{table}
}

\begin{table}[htbp]
    \centering
    \begin{tabular}{ | c | c | c | c | }
    \hline
{Double Pulse Window} & {Rate (Hz)} & {Rate (/day)} & Rate (/year)\\ 
 \hline\hline
1 ms & $8.0\times10^{-3}$& 690 & $2.5\times10^{5}$\\
 \hline
10 $\mu$s & $8.0\times10^{-5}$& $6.9$ & $2.5\times10^{3}$\\
 \hline
1 $\mu$s & $8.0\times10^{-6}$& 0.69 & 250\\
 \hline
250 ns & $2.0\times10^{-6}$& 0.17 & $63$\\
 \hline
\end{tabular}
    \caption{Table of incident thermal neutron singles rate for a 1 ton GAGG detector at sea level. Neutrons are considered thermal if they have $< 10$ keV of energy.}
    \label{tab:GAGGNeutronRate}
\end{table}

Gadolinium is the largest component of GAGG by mass, and it has a relatively high neutron capture cross section. Because of this, it is important to consider the backgrounds that could arise from neutrons arising due to cosmic rays. Using the neutron flux at sea level \cite{ref:neutronFlux}, we calculate that there are 8.0 thermal neutrons per second incident on a 1 ton GAGG detector. Table \ref{tab:GAGGNeutronRate} shows the noise rates of thermal neutrons at sea level if we apply Equation \ref{eq:rateRedux} to several time windows.

\section{Conclusion}\label{sec:Conclusion}
All previous gallium experiments have used radiochemical means to separate out the germanium atoms resulting from solar neutrino interactions for counting. This usually causes the actual measurement of the interaction to be severely out of time with respect to when it took place. To actually see a double pulse, we need to be able to make measurements of the interaction as it is taking place. In this regard, Gadolinium Aluminum Gallium Garnet (GAGG) scintillators are of interest. These crystals are an emerging scintillating material which are fast, have a high light yield, and contain $23\%$ gallium by mass \cite{ref:GAGG}. A detector comprised of crystals like these could look for an electron signal and a simultaneous gamma ray signal of appropriate energy in any nearby segments of the detector. $\beta$-Ga${}_2$O${}_3$, which is $74\%$ gallium by mass, also shows promise as a scintillator candidate, though it is in its infancy as a detector material \cite{ref:Ga2O3}.

A detector made of either of these two materials could look for double-pulse signals within very narrow time windows, on the order of 40 ns. Inside this window, the detector could search for an electron-like signal and a gamma-like signal in the detection volume. By ensuring that the detection happened simultaneously, for the 500 keV and higher gamma rays, or began within 140 ns, $\approx2T_{1/2}$ for the 175 keV gamma ray, the detector would be able to put strict acceptance criteria on gallium solar neutrino interactions and have the potential for reducing background signals by many orders of magnitude. This suppression could be as large as six orders of magnitude when compared with experiments that allow ground-state transitions. This detection method would also allow the detector to take advantage of the relative amplification of the {}$^8$B and {}$^7$Be solar neutrino signals.

\section{Acknowledgements}\label{sec:Acknowledgements}
We would like to thank Vladislav Barinov and Bruce Cleveland who helped us obtain a copy of John Bahcall's original gallium cross section calculation code. 

Funding: This work was supported in part by the NASA NIAC Program [grant number 80NSSC21K1900]

\bibliographystyle{elsarticle-num} 
\bibliography{cas-refs}

\begin{thebibliography}{10}
\expandafter\ifx\csname url\endcsname\relax
  \def\url#1{\texttt{#1}}\fi
\expandafter\ifx\csname urlprefix\endcsname\relax\def\urlprefix{URL }\fi
\expandafter\ifx\csname href\endcsname\relax
  \def\href#1#2{#2} \def\path#1{#1}\fi

\bibitem{ref:GALLEX}
T.~Kirsten, \href{https://doi.org/10.1088/1742-6596/120/5/052013}{Retrospect of {GALLEX}/{GNO}}, Journal of Physics: Conference Series 120~(5) (2008) 052013.
\newblock \href {https://doi.org/10.1088/1742-6596/120/5/052013} {\path{doi:10.1088/1742-6596/120/5/052013}}.
\newline\urlprefix\url{https://doi.org/10.1088/1742-6596/120/5/052013}

\bibitem{ref:SAGE}
J.~N. Abdurashitov, E.~P. Veretenkin, V.~M. Vermul, V.~N. Gavrin, S.~V. Girin, V.~V. Gorbachev, P.~P. Gurkina, G.~T. Zatsepin, T.~V. Ibragimova, A.~V. Kalikhov, T.~V. Knodel, I.~N. Mirmov, N.~G. Khairnasov, A.~A. Shikhin, V.~E. Yants, T.~J. Bowles, W.~A. Teasdale, J.~S. Nico, J.~F. Wilkerson, B.~T. Cleveland, S.~R. Elliott, S.~A. G.~E. Collaboration, \href{https://doi.org/10.1134/1.1506424}{Solar neutrino flux measurements by the soviet-american gallium experiment (sage) for half the 22-year solar cycle}, Journal of Experimental and Theoretical Physics 95~(2) (2002) 181--193.
\newblock \href {https://doi.org/10.1134/1.1506424} {\path{doi:10.1134/1.1506424}}.
\newline\urlprefix\url{https://doi.org/10.1134/1.1506424}

\bibitem{ref:BorexinoBackground}
G.~Bellini, J.~Benziger, D.~Bick, G.~Bonfini, D.~Bravo, M.~Buizza~Avanzini, B.~Caccianiga, L.~Cadonati, F.~Calaprice, P.~Cavalcante, A.~Chavarria, A.~Chepurnov, D.~D'Angelo, S.~Davini, A.~Derbin, A.~Empl, A.~Etenko, K.~Fomenko, D.~Franco, F.~Gabriele, C.~Galbiati, S.~Gazzana, C.~Ghiano, M.~Giammarchi, M.~G\"oger-Neff, A.~Goretti, L.~Grandi, M.~Gromov, C.~Hagner, E.~Hungerford, A.~Ianni, A.~Ianni, V.~Kobychev, D.~Korablev, G.~Korga, D.~Kryn, M.~Laubenstein, T.~Lewke, E.~Litvinovich, B.~Loer, F.~Lombardi, P.~Lombardi, L.~Ludhova, G.~Lukyanchenko, I.~Machulin, S.~Manecki, W.~Maneschg, G.~Manuzio, Q.~Meindl, E.~Meroni, L.~Miramonti, M.~Misiaszek, M.~Montuschi, P.~Mosteiro, V.~Muratova, L.~Oberauer, M.~Obolensky, F.~Ortica, K.~Otis, M.~Pallavicini, L.~Papp, C.~Pena-Garay, L.~Perasso, S.~Perasso, A.~Pocar, G.~Ranucci, A.~Razeto, A.~Re, A.~Romani, N.~Rossi, R.~Saldanha, C.~Salvo, S.~Sch\"onert, H.~Simgen, M.~Skorokhvatov, O.~Smirnov, A.~Sotnikov, S.~Sukhotin, Y.~Suvorov, R.~Tartaglia, G.~Testera, D.~Vignaud, R.~B.
  Vogelaar, F.~von Feilitzsch, J.~Winter, M.~Wojcik, A.~Wright, M.~Wurm, J.~Xu, O.~Zaimidoroga, S.~Zavatarelli, G.~Zuzel, \href{https://link.aps.org/doi/10.1103/PhysRevD.89.112007}{Final results of borexino phase-i on low-energy solar neutrino spectroscopy}, Phys. Rev. D 89 (2014) 112007.
\newblock \href {https://doi.org/10.1103/PhysRevD.89.112007} {\path{doi:10.1103/PhysRevD.89.112007}}.
\newline\urlprefix\url{https://link.aps.org/doi/10.1103/PhysRevD.89.112007}

\bibitem{ref:HuberGAGGDetector}
P.~Huber, \href{https://link.aps.org/doi/10.1103/PhysRevD.107.096011}{Testing the gallium anomaly}, Phys. Rev. D 107 (2023) 096011.
\newblock \href {https://doi.org/10.1103/PhysRevD.107.096011} {\path{doi:10.1103/PhysRevD.107.096011}}.
\newline\urlprefix\url{https://link.aps.org/doi/10.1103/PhysRevD.107.096011}

\bibitem{ref:GAGGContamination}
T.~Omori, T.~Iida, A.~Gando, K.~Hosokawa, K.~Kamada, K.~Mizukoshi, Y.~Shoji, M.~Yoshino, K.-I. Fushimi, H.~Suzuki, K.~Takahashi, \href{https://doi.org/10.1093/ptep/ptae026}{{First Study of the PIKACHU Project: Development and Evaluation of High-Purity Gd3Ga3Al2O12:Ce Crystals for 160Gd Double Beta Decay Search}}, Progress of Theoretical and Experimental Physics 2024~(3) (2024) 033D01.
\newblock \href {http://arxiv.org/abs/https://academic.oup.com/ptep/article-pdf/2024/3/033D01/56875192/ptae026.pdf} {\path{arXiv:https://academic.oup.com/ptep/article-pdf/2024/3/033D01/56875192/ptae026.pdf}}, \href {https://doi.org/10.1093/ptep/ptae026} {\path{doi:10.1093/ptep/ptae026}}.
\newline\urlprefix\url{https://doi.org/10.1093/ptep/ptae026}

\bibitem{ref:MyNIMArticle}
N.~Solomey, J.~Folkerts, H.~Meyer, C.~Gimar, J.~Novak, B.~Doty, T.~English, L.~Buchele, A.~Nelsen, R.~McTaggart, M.~Christl, \href{https://www.sciencedirect.com/science/article/pii/S0168900223000542}{Concept for a space-based near-solar neutrino detector}, Nuclear Instruments and Methods in Physics Research Section A: Accelerators, Spectrometers, Detectors and Associated Equipment 1049 (2023) 168064.
\newblock \href {https://doi.org/https://doi.org/10.1016/j.nima.2023.168064} {\path{doi:https://doi.org/10.1016/j.nima.2023.168064}}.
\newline\urlprefix\url{https://www.sciencedirect.com/science/article/pii/S0168900223000542}

\bibitem{ref:MOON}
R.~Hazama, P.~Doe, H.~Ejiri, S.~Elliott, J.~Engel, M.~Finger, J.~Formaggio, K.~Fushimi, V.~Gehman, A.~Gorin, M.~Greenfield, K.~Ichihara, Y.~Ikegami, H.~Ishii, T.~Itahashi, P.~Kavitov, V.~Kekelidze, K.~Kuroda, V.~Kutsalo, I.~Manouilov, K.~Matsuoka, H.~Nakamura, M.~Nomachi, A.~Para, K.~Rielage, A.~Rjazantsev, R.~Robertson, Y.~Shichijo, T.~Shima, Y.~Shimada, G.~Shirkov, A.~Sissakian, Y.~Sugaya, A.~Titov, V.~Vatulin, O.~Vilches, V.~Voronov, J.~Wilkerson, D.~Will, S.~Yoshida, \href{https://www.sciencedirect.com/science/article/pii/S0920563204005900}{Spectroscopy of low energy solar neutrinos by moon: -mo observatory of neutrinos-}, Nuclear Physics B - Proceedings Supplements 138 (2005) 102--105, proceedings of the Eighth International Workshop on Topics in Astroparticle and Undeground Physics.
\newblock \href {https://doi.org/https://doi.org/10.1016/j.nuclphysbps.2004.11.025} {\path{doi:https://doi.org/10.1016/j.nuclphysbps.2004.11.025}}.
\newline\urlprefix\url{https://www.sciencedirect.com/science/article/pii/S0920563204005900}

\bibitem{ref:A=71}
K.~Abusaleem, B.~Singh, \href{https://www.sciencedirect.com/science/article/pii/S0090375210001213}{Nuclear data sheets for a = 71}, Nuclear Data Sheets 112~(1) (2011) 133--273.
\newblock \href {https://doi.org/https://doi.org/10.1016/j.nds.2010.12.002} {\path{doi:https://doi.org/10.1016/j.nds.2010.12.002}}.
\newline\urlprefix\url{https://www.sciencedirect.com/science/article/pii/S0090375210001213}

\bibitem{ref:GAGG}
Y.~Zhu, S.~Qian, Z.~Wang, H.~Guo, L.~Ma, Z.~Wang, Q.~Wu, \href{https://www.sciencedirect.com/science/article/pii/S0925346720303098}{Scintillation properties of gagg:ce ceramic and single crystal}, Optical Materials 105 (2020) 109964.
\newblock \href {https://doi.org/https://doi.org/10.1016/j.optmat.2020.109964} {\path{doi:https://doi.org/10.1016/j.optmat.2020.109964}}.
\newline\urlprefix\url{https://www.sciencedirect.com/science/article/pii/S0925346720303098}

\bibitem{ref:doty2024gammaraydetectionefficiency}
B.~Doty, N.~Solomey, J.~Folkerts, B.~Hartsock, H.~Meyer, M.~Christl, M.~Rodriguez-Otero, E.~Kuznetsov, \href{https://arxiv.org/abs/2405.04659}{Gamma ray detection efficiency of gagg crystal scintillator using three tagged gamma ray techniques} (2024).
\newblock \href {http://arxiv.org/abs/2405.04659} {\path{arXiv:2405.04659}}.
\newline\urlprefix\url{https://arxiv.org/abs/2405.04659}

\bibitem{ref:Ga2O3}
A.~Datta, R.~Toufanian, W.~Zhang, P.~Halasyamani, S.~Motakef, \href{https://www.sciencedirect.com/science/article/pii/S0925346722011521}{Radiation hard gallium oxide scintillators for high count rate radiation detection}, Optical Materials 134 (2022) 113115.
\newblock \href {https://doi.org/https://doi.org/10.1016/j.optmat.2022.113115} {\path{doi:https://doi.org/10.1016/j.optmat.2022.113115}}.
\newline\urlprefix\url{https://www.sciencedirect.com/science/article/pii/S0925346722011521}

\bibitem{ref:brooksThesis}
B.~B. Hartsock, \href{https://soar.wichita.edu/items/66a66e0e-0e53-44bc-9e96-84925b24c13b}{Advancements in voxelated technology for reactor anti-neutrino and space-based solar neutrino detection}, Master's thesis, Wichita State University (May 2024).
\newline\urlprefix\url{https://soar.wichita.edu/items/66a66e0e-0e53-44bc-9e96-84925b24c13b}

\bibitem{ref:Bahcall_CrossSec}
J.~N. Bahcall, \href{https://doi.org/10.1103\%2Fphysrevc.56.3391}{Gallium solar neutrino experiments: Absorption cross sections, neutrino spectra, and predicted event rates}, Physical Review C 56~(6) (1997) 3391--3409.
\newblock \href {https://doi.org/10.1103/physrevc.56.3391} {\path{doi:10.1103/physrevc.56.3391}}.
\newline\urlprefix\url{https://doi.org/10.1103\%2Fphysrevc.56.3391}

\bibitem{ref:GalliumCrossSec3}
D.~Frekers, H.~Ejiri, H.~Akimune, T.~Adachi, B.~Bilgier, B.~Brown, B.~Cleveland, H.~Fujita, Y.~Fujita, M.~Fujiwara, E.~Ganioğlu, V.~Gavrin, E.-W. Grewe, C.~Guess, M.~Harakeh, K.~Hatanaka, R.~Hodak, M.~Holl, C.~Iwamoto, N.~Khai, H.~Kozer, A.~Lennarz, A.~Okamoto, H.~Okamura, P.~Povinec, P.~Puppe, F.~Šimkovic, G.~Susoy, T.~Suzuki, A.~Tamii, J.~Thies, J.~{Van de Walle}, R.~Zegers, \href{https://www.sciencedirect.com/science/article/pii/S0370269311013153}{The 71ga(3he,t) reaction and the low-energy neutrino response}, Physics Letters B 706~(2) (2011) 134--138.
\newblock \href {https://doi.org/https://doi.org/10.1016/j.physletb.2011.10.061} {\path{doi:https://doi.org/10.1016/j.physletb.2011.10.061}}.
\newline\urlprefix\url{https://www.sciencedirect.com/science/article/pii/S0370269311013153}

\bibitem{ref:GalliumCrossSec2}
V.~Barinov, B.~Cleveland, V.~Gavrin, D.~Gorbunov, T.~Ibragimova, \href{https://doi.org/10.1103\%2Fphysrevd.97.073001}{Revised neutrino-gallium cross section and prospects of {BEST} in resolving the gallium anomaly}, Physical Review D 97~(7) (apr 2018).
\newblock \href {https://doi.org/10.1103/physrevd.97.073001} {\path{doi:10.1103/physrevd.97.073001}}.
\newline\urlprefix\url{https://doi.org/10.1103\%2Fphysrevd.97.073001}

\bibitem{ref:GalliumCrossSec4}
C.~Giunti, Y.~Li, C.~Ternes, Z.~Xin, \href{https://www.sciencedirect.com/science/article/pii/S0370269323003179}{Inspection of the detection cross section dependence of the gallium anomaly}, Physics Letters B 842 (2023) 137983.
\newblock \href {https://doi.org/https://doi.org/10.1016/j.physletb.2023.137983} {\path{doi:https://doi.org/10.1016/j.physletb.2023.137983}}.
\newline\urlprefix\url{https://www.sciencedirect.com/science/article/pii/S0370269323003179}

\bibitem{ref:Bahcall_2005}
J.~Bahcall, \href{http://www.sns.ias.edu/~jnb/}{Software and data for solar neutrino research} (Oct 2005).
\newline\urlprefix\url{http://www.sns.ias.edu/~jnb/}

\bibitem{ref:MARLEY}
S.~Gardiner, \href{https://www.sciencedirect.com/science/article/pii/S0010465521002356}{Simulating low-energy neutrino interactions with marley}, Computer Physics Communications 269 (2021) 108123.
\newblock \href {https://doi.org/https://doi.org/10.1016/j.cpc.2021.108123} {\path{doi:https://doi.org/10.1016/j.cpc.2021.108123}}.
\newline\urlprefix\url{https://www.sciencedirect.com/science/article/pii/S0010465521002356}

\bibitem{ref:muonFlux}
E.~V. Bugaev, A.~Misaki, V.~A. Naumov, T.~S. Sinegovskaya, S.~I. Sinegovsky, N.~Takahashi, \href{http://dx.doi.org/10.1103/PhysRevD.58.054001}{Atmospheric muon flux at sea level, underground, and underwater}, Physical Review D 58~(5) (Jul. 1998).
\newblock \href {https://doi.org/10.1103/physrevd.58.054001} {\path{doi:10.1103/physrevd.58.054001}}.
\newline\urlprefix\url{http://dx.doi.org/10.1103/PhysRevD.58.054001}

\bibitem{ref:neutronFlux}
J.~F. Ziegler, Terrestrial cosmic rays, IBM Journal of Research and Development 40~(1) (1996) 19--39.
\newblock \href {https://doi.org/10.1147/rd.401.0019} {\path{doi:10.1147/rd.401.0019}}.

\end{thebibliography}


\end{document}